# Dynamics and Origin of Comets: New Problems Appeared after the Rosetta Space Mission

V. V. Emel'yanenko

Institute of Astronomy, Russian Academy of Sciences, Moscow, 109017 Russia
e-mail: vvemel@inasan.ru

The data obtained in the recent Rosetta space mission to comet 67P/Churyumov–Gerasimenko have had a profound impact on the understanding of the nature of comets. In addition to revising the notions on the physical properties and structure of comets, this addresses dynamical aspects of the formation of the observed cometary populations (short- and long-period comets, Centaurs, trans-Neptunian objects, and Oort-cloud objects). In the review, we discuss new problems that have appeared in the theory of dynamical evolution and origin of comets due to the Rosetta mission.

INTRODUCTION

The study of comets is a field of particular interest in astronomy, since comets are supposed to preserve the pristine material from which the present bodies of the Solar System were formed. Consequently, the problems of the origin and evolution of comets are considered in close connection with concepts regarding the origin of the Solar System, the formation of planets, and, in the end, the origin of life on the Earth. Together with the use of the best ground-based instruments, many space missions are conducting intensive study of the nature of comets. Among these missions, there were also rather close fly-bys of space probes (within 1000 km of an investigated comet): the Giotto mission accompanied by the Vega-1 and -2 spacecraft to comet Halley (1986) and its extension to Comet Grigg−Skjellerup (1992), the Stardust mission to Comet Wild 2 (2004) and its extension NExT to Comet Tempel 1 (2011), and the Deep Impact mission to Comet Tempel 1 (2005) and its extension EPOXI to Comet Hartley 2 (2010). These space investigations have led to substantial progress in the understanding of the nature of comets. However, the recent Rosetta mission to Comet 67P/Churyumov–Gerasimenko is undoubtedly at a new, previously unachievable, level of cometary studies. For two years, the spacecraft was moving in close vicinity of the comet; and, on February 14, 2015, approaching to 6 km of the comet. During the mission, the Philae probe landed on the surface of the comet. At the end of the mission, on September 30, 2016, the spacecraft gradually approached the comet and bumped into its surface. An extremely wide set of instruments was used in the space experiment, and the data were transmitted up to the moment when the spacecraft collided with the comet.

     Naturally, the results of the space mission Rosetta had a great influence on the views of the nature of comets. This concerns not only the revision of views on the physical properties and structure of comets, but also affects the dynamical aspects of the formation of observable comet populations. There is no doubt that the physics and dynamics of comets are interrelated. It is impossible to understand the data on the physical and chemical characteristics of comets without having information about where they came from and how they reached observed orbits. On the other hand, it is impossible to study the long-term evolution of the orbits of comets without knowing their physical properties. In particular, comets are affected by non-gravitational forces, the magnitude of which depends on many physical characteristics of comets. Analysis of the distribution of the orbits of comets is associated with the effects of observational selection, which require knowledge of how long and in which way the physical activity of comets has changed. However, the main discussion in this article focuses on the new problems that the results of the space mission Rosetta put in the theory of the dynamical evolution and origin of comets.in distant regions of the Solar system.

Since the physical and dynamical lifetime of comets in near-Earth space is rather short, as compared to the age of the Solar System, it is natural to suppose that comets spend most of their time in the distant regions of the Solar System. To determine the structure of these remote sources of comets, having existed for billions of years and continuing to eject comets, and the paths of comets from the outer Solar System to the interior is a key problem of cometary dynamics. Naturally, to address this problem, one should also study the evolution and origin of distant cometary populations. In this context, the dynamics of observed comets are related to more general problems of the Solar System cosmogony and the formation of planets.

POPULATIONS OF COMETS IN THE SOLAR SYSTEM

Along with visible features, especially the presence of a coma in the objects of the inner Solar system, there are two characteristic features in comets that set them apart from the other Solar system bodies. First, comets move along high-eccentricity orbits; and, second, there are often substantial nongravitational effects in their motion. In the mid last century, attempts to explain these features resulted in two remarkable achievements in understanding the nature of comets. It was found that comets with near-parabolic orbits exhibit a concentration of values $w = 1/a$ (where $a$ is the semimajor axis of an orbit) in a range of $w<10^{-4}$ au$^{-1}$. This led J.H. Oort to a concept of a cometary cloud on the periphery of the Solar system (Oort, 1950); this cloud now bears his name. The nongravitational effects were explained by the model, supposing the presence of a solid rotating nucleus composed of frozen volatile compounds, mainly water with admixed dust (Whipple, 1950). In our days, the Oort cloud model and the cometary nucleus model have been substantially modified and improved; however, in general, these achievements provided a basis for the approaches currently developed in the studies of cometary dynamics and physics.

While Oort (1950) had only 14 sufficiently exact orbits with $w<10^{-4}$ au$^{-1}$ at his disposal, now the number of such orbits has increased substantially. For near-parabolic comets with a perihelion distance $q<5$ au, typical changes in the value of $w$ due to the planetary perturbations for one passage through the planetary region considerably exceed $10^{-4}$ au$^{-1}$ (Fernandez,1981; Emel'yanenko, 1992). Consequently, there is no doubt that the majority of observed comets with the Oort-cloud values of $w$ are "new": they enter near-Earth space for the first time after long being in orbits with large perihelion distances. As shown in papers by Duncan et al. (1987), Dones et al. (2004), and Emel'yanenko et al. (2007), the Oort cloud is a natural result of a long dynamical evolution of objects ejected from the planetary region. The main features in the orbit distribution of the Oort cloud weakly depend on dynamical characteristics specific to the bodies at the first stages of the Solar system formation and are mainly determined by the long-term influence of planetary, stellar, and galactic perturbations. Due to the effects of stellar and galactic perturbations, the orbits of most objects are currently far from the planetary region and only some of the objects can pass to near-parabolic orbits. Together with an outer part of the Oort cloud ($a>10^4$ au), which the currently observed new comets come from, there is also an inner part of the Oort cloud ($a$ is between $10^3$ and $10^4$ au), from which the comets may be directly ejected into near-Earth space only after rare visits of stars to this region (Hills, 1981). In the outer part of the Oort cloud, the orbits are isotropically distributed, while prograde orbits noticeably dominate the inner part. Comparison of the model results to the characteristics of the observed flux of new comets allows one to find that, in the present epoch, $\sim 6 \times 10^{11}$ corresponding cometary objects should be in the Oort cloud ($a > 10^3$ au) and approximately half of them are in the outer part ($a > 10^4$ au) (Emel'yanenko et al., 2007).

After introducing the Oort cloud concept, the issue of this formation as a source of short-period comets was actively discussed. Many short-period comets were observed in several apparitions, which stimulated the studies of their dynamical and physical characteristics. It is practically assured now that the physical and dynamical lifetime of these objects is extremely small relative to the Solar system age and, consequently, a source permanently replenishing the family of short-period comets should exist. The origin of Halley-type comets (with the Tisserand parameter $T < 2$) is well explained by a diffusion of semimajor axes of objects belonging to the flux of near-parabolic comets. However, intensive simulations of the dynamical evolution of comets from the Oort cloud showed that the orbit distribution of comets captured from the flux of near-parabolic comets with perihelia located in the inner planetary region does not agree with the orbital distribution of the observed Jupiter-family comets ($T > 2$) (Duncan et al., 1988; Quinn et al., 1990; Bailey, 1992). Because of this, ideas on other sources of the Jupiter-family comets were proposed. Fernandez (1980) supposed that the belt of objects beyond the orbit of Neptune, which was earlier introduced in the models of the Solar system formation (Edgeworth, 1943; Kuiper, 1951), is the main source of short-period comets of the Jupiter family. Searching for such objects met with success in 1992, when the first trans-Neptunian object, 1992 QB1, was discovered. At the initial stages of studies of the trans-Neptunian region, it was the earlier-predicted Kuiper belt, containing the objects on orbits with small eccentricities and inclinations, that seemed to be a major source of the Jupiter-family comets. Levison and Duncan (1997) showed that, due to a weak dynamical instability, some objects of the Kuiper belt may reach the orbit of Neptune, and their further dynamical evolution under the action of planetary perturbations yields short-period comets of the Jupiter family. However, at the present day, when more than two thousand objects beyond the orbit of Neptune have been discovered, it is clear that the structure of the trans-Neptunian zone is significantly more complicated than that expected earlier. Together with the predicted Kuiper belt containing the objects with orbital semimajor axes $a < 50$ AU, there is one more numerous class of objects moving in very elongated orbits. Perihelia of the observed orbits of this class are near the orbit of Neptune and in the Kuiper belt. It is this population of objects, moving in high-eccentricity orbits, (often called the scattered-disk objects) that is a main source of bodies entering the planetary region from the trans-Neptunian zone (Duncan and Levison, 1997; Emel'yanenko et al., 2004). There are also trans-Neptunian objects with large values of the perihelion distance; they have recently induced active discussions on whether one more distant planet may exist. However, at present, these objects are not an important source of bodies coming into the planetary region.

  At the same time, a thorough analysis of transition of trans-Neptunian objects to orbits of short-period comets showed that the orbital distribution of cometary objects in the planetary region is difficult to explain if only one source is supposed to be in the trans-Neptunian zone. This becomes obvious from the analysis of orbits of Centaurs (the objects with perihelia between Jupiter and Neptune and semimajor axes smaller than 1000 AU). A large number of Centaurs moving along retrograde orbits clearly points to their origin in the Oort cloud (Emel'yanenko et al., 2005; Kaib et al., 2009). The Oort cloud objects may reach the outer planetary region and enlarge the Centaur family in two ways: by directly changing the perihelion distances under the stellar- and galactic-perturbation effects and due to the long-term evolution under the action of planetary perturbations through a stage of trans-Neptunian orbits with large eccentricities. Later, most Centaurs are ejected by the planets from the Solar System; and some of them may achieve short-period orbits. In the last case, they mainly form a class of the Jupiter-family comets. According to the estimates by Emel'yanenko et al. (2013), the objects in the scattered disk of the trans-Neptunian zone and those in the Oort cloud almost equally contribute to the population of the Jupiter-family comets.

  Thus, some objects, the perihelia of which are rather close to the planetary region, enter the region of $a < 10^3$ au from the Oort cloud and form the trans-Neptunian class of objects, moving in high-eccentricity orbits; this is an addition to the objects remaining in the trans-Neptunian zone on high-eccentricity orbits from the initial stage of formation of the Solar system. Because of this, there is no sharp boundary between the Oort cloud and the trans-Neptunian zone. The family of trans-Neptunian objects, moving in high-eccentricity orbits, is a mixture of the objects existing here over the Solar System lifetime and the objects visiting the Oort cloud during their dynamical history (Emel'yanenko et al., 2007, 2013). Brasser and Morbidelli (2013) stress that the planetary migration model, known as the Nice model (Levison et al., 2011), provides a natural explanation of the common origin of both the trans-Neptunian and Oort-cloud objects from a single population of planetesimals initially located beyond the orbit of Neptune. The authors show that, due to migration of Uranus and Neptune in the disk of planetesimals, not only these planets could pass to the current orbits from the orbits that are closer to the Sun, but a scattered disk of objects could be formed and a portion of planetesimals could be ejected into the Oort cloud zone. Thus, it has been now realized that both the Jupiter-family comets and the comets from the Oort cloud were formed in a common process in the outer Solar system. This position also finds support in both the analysis of the volatiles composition (A'Hearn et al., 2012) and the dynamical models connecting the orbit distributions of different populations of comets (Emel'yanenko et al., 2007, 2013; Brasser and Morbidelli, 2013), as well as in the study of physical characteristics of comets (Meech, 2017).

However, the Nice model presumes that dynamical instability in the planetary system appears ~0.5 Gyr after the formation of planets, i.e., the Oort cloud started to form rather late. On the one hand, this scenario was supported by the results of the study of the Oort cloud evolution in a cluster of stars, almost definitely surrounding the Sun at the first stages of the Solar system formation (Nordlander et al., 2017). The authors showed that, under the expected parameters of the cluster, an external part of the Oort cloud is almost completely lost. From this viewpoint, the late formation of the Oort cloud after the disappearance of the cluster of stars from the Sun's vicinity is more acceptable for explaining the currently observed structure of the cometary cloud. However, it should be noted that Nordlander et al. (2017) discuss the already formed Oort cloud and ignore the process of its formation.

On the other hand, in the model that assumes a late start of ejection of bodies to the Oort cloud zone, the collisional processing of cometary objects plays a key role. Levison et al. (2008) showed that, to explain the currently observed distribution of bodies in the Solar system with the Nice model, it is necessary to assume the existence of approximately a thousand of Pluto-size bodies in the planetesimal disk located beyond the planetary orbits. These massive bodies should have induced a substantial excitation of the disk, where the spread in the velocity distribution of planetesimals was on average about 0.5−1 km/s (Levison et al., 2011). In the papers by Morbidelli and Rickman (2015) and Rickman et al. (2015), the evolution of such a disk during 0.4 Gyr was considered. The authors showed that the kilometer-size cometary objects should have experienced numerous mutual collisions for the considered time interval. This led the authors to the conclusion that cometary nuclei are not primordial planetesimals but represent the fragments produced in collisions and disruption of much larger parent bodies.

Now, the arguments in favor of a late start of the dynamical instability process in the system of giant planets have substantially weakened. First, according to recent papers, the data on the lunar craters are in better agreement with asteroids than comets as a source of the heavy bombardment of the Moon (Bottke et al., 2012; Rickman et al., 2017). The possibility of explaining the heavy bombardment of the Moon dated at ~3.9 Gyr was one of the main arguments in support of the Nice model of the planetary system formation. However, it is necessary to mention here that the model, explaining the heavy bombardment of the Moon by asteroids, also faces criticism (Minton et al., 2015; Johnson et al., 2016); and, moreover, it has been questioned whether or not a sharp peak in the bombardment of the Moon around 3.9 Gyr is real (Zellner, 2017; Michael et al., 2018). Second, in the papers by Agnor and Lin (2012) and Kaib and Chambers (2016), it is pointed out that the present configuration of the inner planets could hardly survive the late dynamical instability of the giants. However, the strongest arguments against the long-term existence of the primordial disk of planetesimals beyond the orbits of giants, in a region of 15−30 AU, from which, according to the Nice model, the bodies were later ejected to a zone of the scattered disk and the Oort cloud, have appeared due to the results of the Rosetta space mission.

DATA OF THE ROSETTA SPACE MISSION

New insights into the origin and dynamics of comets, following from the data of the *Rosetta* space mission, were given in the paper (Davidsson et al., 2016). A large team of authors, which includes 48 specialists in different fields of astrophysics and dynamics of the Solar system, analyzed the data on comet 67P/Churyumov–Gerasimenko in comparison with numerous ground-based observations and the results of other space missions, as well as laboratory experiments and numerical simulations. A general conclusion is that comets are primordial objects that experienced no strong collisions and thermal changes from the moment of their formation. The main arguments of the cited study are very convincing, and many of them appeared for the first time due to the Rosetta space mission. We discuss them below in more detail.

The most important Ent parameters of comets, which can be determined from ground-based observations with difficulty, are the volume density and porosity of a comet. On the basis of three independent approaches, close estimates of the density were obtained: $535\pm35$ kg/m$^3$ (Preusker и др., 2015), 532±7 kg/m$^3$ (Jorda и др., 2016) и 533±6 kg/m$^3$ (Patzold и др., 2016). The porosity value depends on both the volume density and the composition of the cometary material. For the nucleus of comet 67P, all of the independent estimates yield the porosity value larger than 70%: 72−74% (Patzold et al., 2016), 71±2% (Davidsson et al., 2016), and 71±8% (Fulle et al., 2016a). The values obtained for the nucleus of comet 67P are consistent with the corresponding quantities for comet 9P/Tempel 1 estimated from the results of the Deep Impact space mission: the density is 400 kg/m3 (Richardson et al., 2007), and the porosity is 75−88% (Ernst and Schultz, 2007). Thus, a low density and a high porosity of a nucleus are apparently typical properties of comets.

One more peculiar feature of comet 67P is an extremely low large-scale strength. At the final touchdown site of the Philae lander, the compressive strength reached 2 MPa (Spohn et al., 2015); however, this value characterizes only the small-scale structure of the surface exposed to a strong action of solar radiation (Biele et al., 2015). For the large-scale forms (from 10 m to 1 km), which are more representative of the properties of the nucleus as a whole, the tensile strength is only 3−10 Pa (in any case, not higher than 150 Pa) (Groussin et al., 2015).

The spectral measurements yielded the following results for gas species in the coma of comet 67P: $CO/H_2O$ = 0.13±0.07, $CO_2/H_2O$ =0.08±0.05 (Hässig et al., 2015), $O_2/H_2O$ = 0.0380±0.0085 (Bieler et al., 2015). Moreover, the researchers succeeded in detecting molecular nitrogen with a ratio $N_2/CO$ = (5.70± 0.66)·10$^{-3}$ (Rubin et al., 2015) and argon with a relative content at $^{36}Ar/H_2O$ =(0.1-2.3)·10$^{-5}$ (Balsiger et al., 2015). The presence of extremely volatile ices of CO, $O_2$, $N_2$, and Ar suggests that the temperature inside the nucleus of comet 67P has never been high. According to Davidsson et al. (2016), this temperature cannot exceed 90 K, if the supervolatiles are inside amorphous water ice, and 40 K, if the cometary water ice is in clathrates or crystalline state (Mousis et al., 2016), or even 20 K, if supervolatiles are not frozen into water ice (Luspay-Kuti et al., 2015; Fulle et al., 2016).

These properties, together with the other arguments (the absence of so-called aqueous alterations and the presence of large-scale smooth linear structures on the surface) led Davidsson et al. (2016) to the conclusion that comet 67P is not a collisional rubble pile formed in the aftermath of disruptions of much larger parent bodies. Trying to remain within the Nice model, the authors proposed their formation scheme for small bodies in the outer Solar System. The suggested scenario assumes that the region, extending from 15 to 30 AU, was rich in solid material with a mass of 15 Earth masses (this is the smallest mass consistent with the Nice model) mostly in the form of micrometer-size particles. Due to collisions in a gas environment, these particles can coagulate to form centimeter-size porous pebbles (see, e.g., Blum and Wurm, 2008; Ricci et al., 2010; Zsom et al., 2010; Birnstiel et al., 2012). Later, during the first ~100 kyr, the action of the streaming instability mechanism (see, e.g., Youdin and Goodman, 2005; Johansen et al., 2007; Wahlberg Jansson and Johansen, 2014) resulted in gathering these bodies into trans-Neptunian objects with sizes of 50−400 km. According to this concept (Davisson et al., 2016), comets were formed from the material remaining in the disk (~13−25% of the initial mass of the solid material) after the formation of trans-Neptunian objects. Comets were formed in collisions of pebbles through the hierarchic agglomeration mechanism proposed by Weidenschilling (1997), and the formation process was very slow. Kilometer-size comets were growing for ~3.5 Myr, till the gaseous disk finally dissipated. Further, during ~25 Myr, comets with nuclei ~40 km in size could be formed in the absence of gas species. In this period, the relative velocities of cometary bodies were ~40 m/s. It is estimated that such a disk, which was composed of large trans-Neptunian objects and relatively small comets, existed for ~400 Myr. In this period, ~350 objects grew to near-Pluto sizes. Later, according to the Nice model, catastrophic changes in the distribution of these bodies occurred, which were related to the migration of Neptune and Uranus.

Such a slow growth of comets makes it possible to overcome the problem connected with the heating of large bodies caused by decay of a short-lived isotope 26Al. Moreover, in the model by Davisson et al. (2016), the collision frequency of comets is substantially lower than that in the estimates by Morbidelli and Rickman (2015) and Rickman et al. (2015). The authors assert that many 67P-type comets could avoid disruption in collisions during the considered 400 Myr.

However, this scenario of the origin and evolution of comets met serious objections, and from opposite positions. On the one part, Jutzi et al. (2017) note that the approach by Davidsson et al.

(2016) implies a small mass of the primordial disk and, consequently, a critically small number of objects in the present-day scattered disk. Moreover, these authors present new arguments that comets should have experienced collisions even in the evolution during the last 4 Gyr, if they were in the scattered disk. Though a significant fraction of comets could avoid disruptive collisions, the number of collisions, leading to substantial changes in shape, is very large. However, according to the estimates of the authors of this concept, such collisions do not yield considerable changes in the properties of cometary nuclei (the porosity, the presence of volatiles).

However, as stressed by Fulle et al. (2016) and Fulle and Blum (2017), both the last statement that comet 67P can be a product of collisional evolution and even the scenario by Davisson et al. (2016) do not agree with the observational data of the Rosetta space mission. The Grain Impact Analyzer and Dust Accumulator (GIADA) detected particles of two types in the coma of this comet: densely packed particles with sizes from 0.03 to 1 mm and fluffy aggregates with sizes from 0.2 to 2.5 mm (Fulle et al., 2015). The authors connect the first type with the above-discussed pebbles, the microporosity of which is ~50%. The second type of particle is not numerous relative to the first (less than 15% of the total number of particles (Fulle et al., 2015)). At the same time, one fluffy particle came into the field of view of the high-resolution Micro-Imaging Dust Analysis System (MIDAS). The analysis of its structure showed that the microporosity of such fluffy, so-called fractal, particles exceeds 98.7% (Mannel et al., 2016). These particles are associated with the pristine material that existed in the protoplanetary nebula.

Fractal structures are much more fragile than pebbles. All theoretical and experimental estimates show that such structures could survive in collisions only with relative velocities less than 1 m/s (Blum et al., 2000; Weidling et al., 2009; Guttler et al., 2010; Whizin et al., 2017). This suggests that fractal particles represent the primitive protosolar component surviving in voids between pebbles in the course of accretion of comet 67P, which proceeded with extremely low velocities (Fulle et al., 2016). Fulle et al. (2016a) and Fulle and Blum (2017) suppose that, for comets with such properties, the most acceptable formation mechanism is a gentle gravitational collapse of a cloud of pebbles and fractal particles; this collapse is caused by the streaming instability in the protoplanetary nebula. Further catastrophic collisional processing of comets (Morbidelli and Rickman, 2015; Rickman et al., 2015; Jutzi and Benz, 2017; Jutzi et al., 2017) is also implausible, because the shock pressure required to disintegrate a nucleus is substantially higher than the strength of fractal aggregates. Consequently, such fluffy structures will not survive propagation of a shock wave that should inevitably pass through the whole nucleus to destroy it.

## CONCLUSIONS

Though the viewpoint that the primordial cometary material could survive the collisional processing is still upheld (Morbidelli and Rickman, 2015; Jutzi et al., 2017), the abovementioned arguments against this position seem to be weighty. Foreseeing the appearance of new data that do not support the collisional processing scenario, Morbidelli and Rickman (2015) pointed out that, in such a case, their model linking the formation of the observed scattered disk and the Oort cloud with migration of planets will require considerable modification. According to their opinion, the problem can be solved within the Nice model either at the early beginning of the period of dynamical instability of planets or in a dramatic drop of the size distribution function for cometary bodies of subkilometer size. To find the agreement between these suppositions and the data of the Rosetta space mission is not perceived now as simple and the Nice model does not sound now as natural as it was before the start of this mission. Because of this, alternative theories of the origin and dynamical evolution should be considered.

Further progress toward the solution of this problem is seen in realization of new space missions to comets. The Rosetta mission yielded a wealth of scientific data, which are based on the analysis of the cometary material performed directly in the coma and on the surface of comet 67P. However, the issues on the internal cometary structure actually representing the conditions, under which the objects were formed, still excite intense debate. There are conflicting opinions on the homogeneity degree of the cometary nucleus structure (Vincent et al., 2015; Kofman et al., 2015; Patzold et al., 2016). What is the state of the primordial material in a cometary nucleus? Why does the ratio D/H = (5.3±0.7) × 10$^{-4}$ for water in comet 67P (Altwegg et al., 2015) substantially exceed the values determined earlier for the other short-period comets? How real is a global layered structure of comets (Penasa et al., 2017; Thomas et al., 2015; Fulle et al., 2016)? Do the structures on the surface of comet 67P detected by the panoramic camera onboard the Philae probe actually represent agglomerates of primordial pebbles (Poulet et al., 2016; Fulle et al., 2016a)? In which way does a high ratio of the refractory component and ices in comet 67P (according to different estimates, it ranges from 4±2 (Rotundi et al., 2015) to 8.5 (Fulle et al., 2016a)) agree with the composition of a protoplanetary nebula in the formation region of comets? In which way do the surface features of the comet change with time (Vincent et al., 2017) and what are the final stages of the cometary evolution? Were two parts of the nucleus of comet 67P formed independently at early stages of the Solar System formation (Jutzi and Asphaug, 2015; Fulle and Blum, 2017) or could the collisional processing in the later epoch (Schwartz et al., 2018) lead to the observed shape of a nucleus of the comet? Why do comets exhibit such a various relative content of volatiles (A'Hearn et al., 2012)? Substantial advances in solving these and other problems can be achieved only after extracting the material from deep inside a comet and returning this material to laboratories on the Earth. In this respect, what is encouraging is that NASA decided to consider a new space mission to comet 67P/Churyumov–Gerasimenko, which includes the cometary material return to the Earth, as one of two finalists for the final selection in 2019 connected with the scientific program for launch in 2025.

ACKNOWLEDGMENTS

This work was supported by the Russian Science Foundation (project no. 17-12-01441).

REFERENCES

*Agnor C.B., Lin D.N.C.* On the migration of Jupiter and Saturn: constraints from linear models of secular resonant coupling with the terrestrial planets // Astrophys. J. 2012. V.745. Article id. 143. 20 p.

*A'Hearn M.F.*, *Feaga L.M.*, Keller H.U., *Kawakita H.*, Hampton D.L., Kissel J., Klaasen K.P., McFadden L.A., *Meech K.J.*, Schultz P.H., *Sunshine J.M.*, *Thomas P.C.*, *Veverka J*, Yeomans D.K., *Besse S.*, *Bodewits D.*, *Farnham T.L.*, *Groussin O.*, *Kelley M.S.*, *Lisse C.M.*, *Merlin F.*, Protopapa S., *Wellnitz D.D.* Cometary volatiles and the origin of comets // Astrophys. J. 2012. V. 758. Article id. 29. 8 p.

Altwegg K., Balsiger H., Bar-Nun A., Berthelier J.J., Bieler A., Bochsler P., Briois C., Calmonte U., Combi M., De Keyser J., Eberhardt P., Fiethe B., Fuselier S., Gasc S., Gombosi T.I., Hansen K.C., Hassig M., Jackel A., Kopp E., Korth A., LeRoy L., Mall U., Marty B., Mousis O., Neefs E., Owen T., Reme H., Rubin M., Sémon T., Tzou C.Y., Waite H., Wurz P. 67P/Churyumov-Gerasimenko, a Jupiter family comet with a high D/H ratio. Science. 2015. V. 347. Article id. 1261952.

*Bailey M.E.* Origin of short-period comets // Celestial Mechanics and Dynamical Astronomy. 1992. V. 54. P. 49-61.

*Balsiger H.*, *Altwegg K.*, *Bar-Nun A.*, *Berthelier J.J.*, *Bieler A.*, *Bochsler P.*, *Briois C.*, *Calmonte U.*, *Combi M.*, *De Keyser J.*, *Eberhardt P.*, *Fiethe B.*, *Fuselier S.A.*, *Gasc S.*, *Gombosi T.I.*, *Hansen K.C.*, *Hässig M.*, *Jäckel A.*, *Kopp E.*, *Korth A.*, *Le Roy L.*, *Mall U.*, *Marty B.*, *Mousis O.*, *Owen T.*, *Rème H.*, *Rubin M.*, *Sémon T.*, *Tzou C.Y.*, *Waite J.H.*, *Wurz P*. Detection of argon in the coma of comet 67P/Churyumov-Gerasimenko // Sci. Adv. 2015. 1. e1500377.

*Biele J.*, *Ulamec S.*, *Maibaum M.*, *Roll R.*, *Witte L.*, *Jurado E.*, *Muñoz P.*, *Arnold W.*, *Auster H.-U.*, *Casas C.*, *Faber, C.*, *Fantinati C.*, *Finke F.*, *Fischer H.-H.*, *Geurts K.*, *Güttler C.*, *Heinisch P.*, *Herique A.*, *Hviid S.*, *Kargl G.*, *Knapmeyer M.*, *Knollenberg J.*, *Kofman W.*, *Kömle N.*, Kührt E., *Lommatsch V.*, *Mottola S.*, *Pardo de Santayana R.*, *Remetean E.*, *Scholten F.*, *Seidensticker K.J.*, *Sierks H.*, *Spohn T.* The landing(s) of Philae and inferences about comet surface mechanical properties // Science. 2015. V. 349. aaa9816.

*Bieler A.*, *Altwegg K.*, *Balsiger H.*, *Bar-Nun A.*, *Berthelier J.-J.*, *Bochsler P.*, *Briois C.*, *Calmonte U.*, *Combi M.*, *De Keyser J.*, *Van Dishoeck E.F.*, *Fiethe B.*, *Fuselier S.A.*, *Gasc S. Gombosi T.I.*, Hansen K.C., *Hässig M.*, *Jäckel A.*, *Kopp E.*, *Korth A.*, *Le Roy L.*, *Mall U.*, *Maggiolo R.*,

*Marty B.*, Mousis O., *Owen T.*, *Rème H.*, *Rubin M.*, *Sémon T.*, *Tzou C.-Y.*, *Waite J. H.*, *Walsh C.*, *Wurz P.* Abundant molecular oxygen in the coma of comet 67P/Churyumov-Gerasimenko // Nature. 2015. V. 526. P. 678-681.

*Birnstiel T., Klahr H., Ercolano B.* A simple model for the evolution of the dust population in protoplanetary disks // Astron. Astrophys. 2012. V. 539, Article id. A148.

*Blum J., Wurm G., Kempf S., Poppe T., Klahr H., Kozasa T., Rott M., Henning T., Dorschner J., Schräpler R., Keller H.U., Markiewicz W.J., Mann I., Gustafson B.A., Giovane F., Neuhaus D., Fechtig H., Grün E., Feuerbacher B., Kochan H., Ratke L., El Goresy A., Morfill G.,* Weidenschilling  *S.J., Schwehm G., Metzler K., Ip W.-H.* Growth and form of planetary seedlings: results from a microgravity aggregation experiment // Physical Review Letters. 2000. V. 85. P. 2426-2429.

*Blum J., Wurm G.* The growth mechanisms of macroscopic bodies in protoplanetary disks // Annual Review of Astronomy and Astrophysics. 2008. V. 46. P. 21-56.

*Bottke W.F., Vokrouhlický D., Minton D., Nesvorný D., Morbidelli A., Brasser R., Simonson B., Levison, H. F.* An Archaean heavy bombardment from a destabilized extension of the asteroid belt // Nature. 2012. V. 485. P. 78-81.

*Brasser R., Morbidelli A.* Oort cloud and Scattered Disc formation during a late dynamical instability in the Solar System // Icarus. 2013. V. 225. P. 40-49.

*Davidsson B.J.R., Sierks H., Güttler C., Marzari F., Pajola M., Rickman H., A'Hearn M.F.,* Auger A.-T., *El-Maarry M.R., Fornasier S., Gutiérrez P. J., Keller H.U., Massironi M., Snodgrass C., Vincent J.-B., Barbieri C., Lamy P.L., Rodrigo R., Koschny D., Barucci M.A., Bertaux J.-L., Bertini I., Cremonese G., Da Deppo V., Debei S., De Cecco M., Feller C., Fulle M., Groussin O., Hviid S.F., Höfner S., Ip W.-H., Jorda L., Knollenberg J., Kovacs G., Kramm J.-R., Kührt E., Küppers M., La Forgia F., Lara L.M., Lazzarin M., Lopez Moreno J.J., Moissl-Fraund R., Mottola S., Naletto G., Oklay N., Thomas N., Tubiana C.* The primordial nucleus of comet 67P/Churyumov-Gerasimenko // Astron. and Astrophys. 2016. V. 592. Article id.A63. 30 p.

*Dones L., Weissman P.R., Levison H.F., Duncan M.J.* Oort cloud formation and dynamics // Comets II / Eds. Festou M.C., Keller H.U., Weaver H.A., Tucson: Univ. of Arizona Press, 2004. P. 153-174.

*Duncan M., Quinn T., Tremaine S.* The formation and extent of the solar system comet cloud // Astron. J. 1987. V. 94. P. 1330–1338.

*Duncan M., Quinn T., Tremaine S.* The origin of short-period comets // Astrophys. J. 1988. V. 328. P. L69-L73.

*Duncan M.J., Levison H.F.* A disk of scattered icy objects and the origin of Jupiter-family comets // Science. 1997. V. 276, P. 1670–1672.

*Edgeworth K.E.* The evolution of our planetary system // Monthly Notices of the Royal Astronomical Society. 1943. V. 109. P. 600-609.

*Emel'yanenko V.V.* Dynamics of periodic comets and meteor streams // Celestial Mechanics and Dynamical Astronomy. 1992. V. 54. P. 91-110.

*Emel'yanenko V.V., Asher D.J., Bailey M.E.* High-eccentricity trans-Neptunian objects as a source of Jupiter-family comets // Monthly Notices of the Royal Astronomical Society. 2004. V. 350. P. 161-166.

*Emel'yanenko V.V., Asher D.J., Bailey M.E.* Centaurs from the Oort cloud and the origin of Jupiter-family comets // Monthly Notices of the Royal Astronomical Society. 2005. V. 361. P. 1345-1351.

*Emel'yanenko V.V., Asher D.J., Bailey M.E.* The fundamental role of the Oort cloud in determining the flux of comets through the planetary system // Monthly Notices of the Royal Astronomical Society. 2007. V. 381. P. 779-789.

*Emel'yanenko V.V., Asher D.J., Bailey M.E.* A Model for the Common Origin of Jupiter Family and Halley Type Comets // Earth, Moon and Planets. 2013. V. 110. P. 105-130.

*Ernst C.M., Schultz P.H.* Evolution of the Deep Impact flash: Implications for the nucleus surface based on laboratory experiments // Icarus. 2007. V. 190. P. 334-344.

*Fernandez J.A.* On the existence of a comet belt beyond Neptune. Monthly Notices of the Royal Astronomical Society. 1980. V. 192. P. 481–491.

*Fernandez J.A.* New and evolved comets in the solar system // Astron. and Astrophys. 1981. V. 96. P. 26-35.

*Fulle M., Della Corte V., Rotundi A., Weissman P., Juhasz A., Szego K., Sordini R., Ferrari M., Ivanovski S., Lucarelli F., Accolla M., Merouane S., Zakharov V., Mazzotta Epifani E., López-Moreno J.J., Rodríguez J., Colangeli L., Palumbo P., Grün E., Hilchenbach M., Bussoletti E., Esposito F., Green S.F., Lamy P.L., McDonnell J.A.M., Mennella V., Molina A., Morales R., Moreno F., Ortiz, J.L., Palomba E., Rodrigo R., Zarnecki J.C., Cosi M., Giovane F., Gustafson B., Herranz M.L., Jerónimo J.M., Leese M.R., López-Jiménez A.C., Altobelli N.* Density and charge of pristine fluffy particles from Comet 67P/Churyumov-Gerasimenko // Astrophys. J. Letters. 2015. V. 802. Article id. L12. 5 p.

*Fulle M., Altobelli N., Buratti B., Choukroun M., Fulchignoni M., Grün E., Taylor M.G.G.T., Weissman P.* Unexpected and significant findings in comet 67P/Churyumov-Gerasimenko: an interdisciplinary view // Monthly Notices of the Royal Astronomical Society. 2016. V. 462. P. S2-S8.

*Fulle M., Della Corte V., Rotundi A., Rietmeijer F.J.M., Green S.F., Weissman P., Accolla M., Colangeli L., Ferrari M., Ivanovski S., Lopez-Moreno J.J., Mazzotta Epifani E., Morales R., Ortiz J.L., Palomba E., Palumbo P., Rodriguez J., Sordini R., Zakharov V.* Comet 67P/Churyumov-Gerasimenko preserved the pebbles that formed planetesimals // Monthly Notices of the Royal Astronomical Society. 2016a. V. 462. P. S132-S137.

*Fulle M., Blum J.* Fractal dust constrains the collisional history of comets // Monthly Notices of the Royal Astronomical Society. 2017. V. 469. P. S39-S44.

*Groussin O., Jorda L., Auger A.-T., Kührt E., Gaskell R., Capanna C., Scholten F., Preusker F., Lamy P., Hviid S., Knollenberg J., Keller U., Huettig C., Sierks H., Barbieri C., Rodrigo R., Koschny D., Rickman H., A'Hearn M.F., Agarwal J., Barucci M. A., Bertaux J.-L., Bertini I., Boudreault S., Cremonese G., Da Deppo V., Davidsson B., Debei S., De Cecco M., El-Maarry M.R., Fornasier S., Fulle M., Gutiérrez P. J., Güttler C., Ip W.-H., Kramm J.-R., Küppers M., Lazzarin M., Lara L.M., Lopez Moreno J. J., Marchi S., Marzari F., Massironi M., Michalik H., Naletto G., Oklay N., Pommerol A., Pajola M., Thomas N., Toth I., Tubiana C., Vincent J.-B.* Gravitational slopes, geomorphology, and material strengths of the nucleus of comet 67P/Churyumov-Gerasimenko from OSIRIS observations // Astron. and Astrophys. 2015. V. 583. Article id. A32. 12 p.

*Güttler C., Blum J., Zsom A., Ormel C.W., Dullemond C.P.* The outcome of protoplanetary dust growth: pebbles, boulders, or planetesimals? I. Mapping the zoo of laboratory collision experiments // Astron. and Astrophys. 2010. V. 513. Article id. A56. 16 p.

*Hässig M., Altwegg K., Balsiger H., Bar-Nun A., Berthelier J.J., Bieler A., Bochsler P., Briois C., Calmonte U., Combi M., De Keyser J., Eberhardt P., Fiethe B., Fuselier S.A., Galand M., Gasc S., Gombosi T.I., Hansen K.C., Jäckel A., Keller H.U., Kopp E., Korth A., Kührt E., Le Roy L., Mall U., Marty B., Mousis O., Neefs E., Owen T., Rème H., Rubin M., Sémon T., Tornow C., Tzou C.-Y., Waite J.H. Wurz P.* Time variability and heterogeneity in the coma of 67P/Churyumov-Gerasimenko // Science. 2015. V. 347. aaa0276.

*Hills J.G.* Comet showers and the steady-state infall of comets from the Oort cloud //Astron. J. 1981. V. 86. P. 1730–1740.

*Johansen A., Oishi J.S., Mac Low M.-M., Klahr H., Henning T.,Youdin A.* Rapid planetesimal formation in turbulent circumstellar disks // Nature. 2007. V. 448. P. 1022-1025.

*Johnson B.C., Collins G.S., Minton D.A., Bowling T.J., Simonson B.M., Zuber M.T.* Spherule layers, crater scaling laws, and the population of ancient terrestrial impactors // Icarus. 2016. V. 271. P. 350-359.

*Jorda L., Gaskell R., Capanna C., Hviid S., Lamy P., Ďurech J., Faury G., Groussin O., Gutiérrez P., Jackman C., Keihm S.J., Keller H.U., Knollenberg J., Kührt E., Marchi S., Mottola S., Palmer E., Schloerb F.P., Sierks H., Vincent J.-B., A'Hearn M.F., Barbieri C., Rodrigo R., Koschny D., Rickman H., Barucci M.A., Bertaux J.L., Bertini I., Cremonese G., Da Deppo V., Davidsson B., Debei S., De Cecco M., Fornasier S., Fulle M., Güttler C., Ip W.-H., Kramm J.R., Küppers M., Lara L.M., Lazzarin M., Lopez Moreno J.J., Marzari F., Naletto G., Oklay N.,*

*Thomas N.*, *Tubiana C.*, *Wenzel K.-P.* The global shape, density and rotation of Comet 67P/Churyumov-Gerasimenko from preperihelion Rosetta/OSIRIS observations // Icarus. 2016. V. 277. P. 257-278.

*Jutzi M.*, *Asphaug E.* The shape and structure of cometary nuclei as a result of low-velocity accretion // Science. 2015. V. 348. P. 1355-1358.

*Jutzi M.*, *Benz W.*, *Toliou A.*, *Morbidelli A.*, *Brasser R.* How primordial is the structure of comet 67P? Combined collisional and dynamical models suggest a late formation // Astron. and Astrophys. 2017. V. 597. Article id. A61. 13 p.

*Jutzi M.*, *Benz W.* Formation of bi-lobed shapes by sub-catastrophic collisions. A late origin of comet 67P's structure // Astron. and Astrophys . 2017. V. 597. Article id. A62. 10 p.

*Kaib N.A.*, *Becker A.C.*, *Jones R.L.*, *Puckett A.W.*, *Bizyaev D.*, *Dilday B.*, *Frieman J.A.*, Oravetz Pan K., *Quinn T.*, *Schneider D.P.*, *Watters S.* 2006 $SQ_{372}$: A likely long-period comet from the inner Oort cloud // Astrophys. J. 2009. V. 695. P. 268-275.

*Kaib N.A., Chambers J.E.* The fragility of the terrestrial planets during a giant-planet instability // Monthly Notices of the Royal Astronomical Society. 2016.V. 455. P. 3561-3569.

*Kofman W.*, *Herique A.*, *Barbin Y.*, *Barriot J.-P.*, *Ciarletti V.*, *Clifford S.*, *Edenhofer P.*, Elachi C., *Eyraud C.*, *Goutail J.-P.*, *Heggy E.*, *Jorda L.*, *Lasue J.*, *Levasseur-Regourd A.-C.*, *Nielsen E.*, *Pasquero P.*, *Preusker F.*, *Puget P.*, *Plettemeier D.*, *Rogez Y.*, *Sierks H.*, *Statz C.*, *Svedhem H.*, *Williams I.*, *Zine S.*, *Van Zyl J.* Properties of the 67P/Churyumov-Gerasimenko interior revealed by CONSERT radar // Science. 2015. V. 349. aab0639.

*Kuiper G.P.* On the origin of the Solar system // Astrophysics: A Topical Symposium. New York: McGraw-Hill, 1951. P. 357-424.

*Levison H.F., Duncan M.J.* From the Kuiper Belt to Jupiter-family comets: the spatial distribution of ecliptic comets // Icarus. 1997. V. 127. P. 13–32.

Levison H.F., Morbidelli A., Vanlaerhoven C., Gomes R., Tsiganis K. Origin of the structure of the Kuiper belt during a dynamical instability in the orbits of Uranus and Neptune // Icarus. 2008. V. 196. P. 258-273.

Levison H.F., Morbidelli A., Tsiganis K., Nesvorný D., Gomes R. Late orbital instabilities in the outer planets induced by interaction with a self-gravitating planetesimal disk // Astron. J. 2011. V. 142. Id. 152. 11p.

*Luspay-Kuti A.*, *Hässig M.*, *Fuselier S.A.*, *Mandt K.E.*, *Altwegg K.*, *Balsiger H.*, *Gasc S.*, *Jäckel A.*, *Le Roy L.*, *Rubin M.*, *Tzou C.-Y.*, *Wurz P.*, *Mousis O.*, *Dhooghe F.*, *Berthelier J.J.*, *Fiethe B.*, *Gombosi T.I.*, *Mall U.* Composition-dependent outgassing of comet 67P/Churyumov-Gerasimenko from ROSINA/DFMS. Implications for nucleus heterogeneity? // Astron. and Astrophys. 2015. Vol. 583. Article id. A4. 8 p.

*Mannel T.*, *Bentley M.S.*, *Schmied R.*, *Jeszenszky H.*, *Levasseur-Regourd A.C.*, *Romstedt J.*, *Torkar K.* Fractal cometary dust - a window into the early Solar system // Monthly Notices of the Royal Astronomical Society. 2016. V. 462. P .S304-S311.

*Meech K.J.* Setting the scene: what did we know before Rosetta? // Philosophical Transactions of the Royal Society A. 2017. V. 375. Id. 20160247.

*Michael G.*, *Basilevsky A.*, *Neukum G.* On the history of the early meteoritic bombardment of the Moon: was there a terminal lunar cataclysm? // Icarus. 2018. V. 302. P. 80-103.

*Minton D.A.*, *Richardson J.E.*, *Fassett C.I.* Re-examining the main asteroid belt as the primary source of ancient lunar craters // Icarus. 2015. V. 247. P. 172-190.

*Morbidelli A.*, *Rickman H.* Comets as collisional fragments of a primordial planetesimal disk // Astron. and Astrophys. 2015. V. 583 Article id. A43. 9 p.

*Mousis O.*, *Lunine J.I.*, *Luspay-Kuti A.*, *Guillot T.*, *Marty B.*, *Ali-Dib M.*, *Wurz P.*, *Altwegg K.*, *Bieler A.*, *Hässig M.*, *Rubin M.*, *Vernazza P.*, *Waite J.H.* A protosolar nebula origin for the ices agglomerated by Comet 67P/Churyumov-Gerasimenko // Astrophys. J. Letters. 2016. V. 819. Article id. L33. 5 p.

*Nordlander T.*, *Rickman H.*, *Gustafsson B.* The destruction of an Oort Cloud in a rich stellar cluster // Astron. and Astrophys. 2017. V. 603. Article id. A112. 18 p.

*Oort J.H.* The structure of the cloud of comets surrounding the Solar System and a hypothesis concerning its origin // Bull. Astron. Inst. Neth. 1950. V. 11. P. 91–110.

*Pätzold M.*, *Andert T.*, *Hahn M.*, *Asmar S.W.*, *Barriot J.-P.*, *Bird M.K.*, *Häusler B.*, *Peter K.*, *Tellmann S.*, *Grün E.*, *Weissman P.R.*, *Sierks H.*, *Jorda L.*, *Gaskell R.*, *Preusker F.*, *Scholten F.* A homogeneous nucleus for comet 67P/Churyumov-Gerasimenko from its gravity field // Nature. 2016. V. 530. P. 63-65.

*Penasa L.*, *Massironi M.*, *Naletto G.*, *Simioni E.*, *Ferrari S.*, *Pajola M.*, *Lucchetti A.*, *Preusker F.*, *Scholten F.*, *Jorda L.*, *Gaskell R.*, *Ferri F.*, *Marzari F.*, *Davidsson B.*, *Mottola S.*, *Sierks H.*; Barbieri C., *Lamy P.L.*, *Rodrigo R.*, *Koschny D.*, *Rickman H.*, *Keller H.U.*, *Agarwal J.*, *A'Hearn M.F.*, Barucci M.A., *Bertaux J.L.*, *Bertini I.*, *Cremonese G.*, *Da Deppo V.*, *Debei S.*, *De Cecco M.*, *Deller J.*, Feller C., *Fornasier S.*, *Frattin E.*, *Fulle M.*, *Groussin O.*, *Gutierrez P.J.*; *Güttler C.*, *Hofmann M.*, Hviid S.F., *Ip W.H.*, *Knollenberg J.*, *Kramm J. R.*, *Kührt E.*, *Küppers M.*, *La Forgia F.*, *Lara L.M.*, *Lazzarin M.*, *Lee J.-C.*, *Lopez Moreno J.J.*, *Oklay N.*, *Shi X.*, *Thomas N.*, *Tubiana C.*, *Vincent J.B.* A three dimensional modelling of the layered structure of comet 67P/Churyumov-Gerasimenko // Monthly Notices of the Royal Astronomical Society. 2017. V. 469. P. S741-S754.

*Poulet F.*, *Lucchetti A.*, *Bibring J.-P.*, *Carter J.*, *Gondet B.*, *Jorda L.*, *Langevin Y.*, *Pilorget C.*, *Capanna C.*, *Cremonese G.* Origin of the local structures at the Philae landing site and possible implications on the formation and evolution of 67P/Churyumov-Gerasimenko // Monthly Notices of the Royal Astronomical Society. 2016. V. 462. P. S23-S32.

*Preusker F.*, *Scholten F.*, *Matz K.-D.*, *Roatsch T.*, *Willner K.*, *Hviid S.F.*, *Knollenberg J.*, *Jorda L.*, *Gutiérrez P.J.*, *Kührt E.*, *Mottola S.*, *A'Hearn M.F.*, *Thomas N.*, *Sierks H.*, *Barbieri C.*, *Lamy P.*, *Rodrigo R.*, *Koschny D.*, *Rickman H.*, *Keller H.U.*, *Agarwal J.*, *Barucci M.A.*; *Bertaux J.-L.*, Bertini I., *Cremonese G.*, *Da Deppo V.*, *Davidsson B.*, *Debei S.*, *De Cecco M.*, *Fornasier S.*, *Fulle M.*, *Groussin O.*, *Güttler C.*, *Ip W.-H.*, *Kramm J. R.*, *Küppers M.*, *Lara L.M. Lazzarin M.*, Lopez Moreno J. J., *Marzari F.*, *Michalik H.*, *Naletto G.*, *Oklay N.*, *Tubiana C.*, *Vincent J.-B.* Shape model, reference system definition, and cartographic mapping standards for comet 67P/Churyumov-Gerasimenko - Stereo-photogrammetric analysis of Rosetta/OSIRIS image data // Astron. and Astrophys. 2015. V. 583. Article id. A33. 19 p.

*Quinn T., Tremaine S., Duncan M.* Planetary perturbations and the origin of short-period comets // Astrophys. J. 1990. V. 355. P. 667-679.

*Ricci L.*, *Testi L.*, *Natta A.*, *Neri R.*, *Cabrit S.*, *Herczeg G.J.* Dust properties of protoplanetary disks in the Taurus-Auriga star forming region from millimeter wavelengths // Astron. and Astrophys. 2010. V. 512. Article id.A15, 15 p.

*Richardson J.E.*, *Melosh H.J.*, *Lisse C.M.*, *Carcich B.* A ballistics analysis of the Deep Impact ejecta plume: Determining Comet Tempel 1's gravity, mass, and density // Icarus. 2007. V. 190. P. 357-390.

*Rickman H.*, *Marchi S.*, *A'Hearn M.F.*, *Barbieri C.*, *El-Maarry M.R.*, *Güttler C.*, *Ip, W.-H.*, *Keller H.U.*, *Lamy P.*, *Marzari F.*, *Massironi M.*, *Naletto G.*, *Pajola M.*, *Sierks H.*, *Koschny D.*, *Rodrigo R.*, *Barucci M.A.*, *Bertaux J.-L.*, *Bertini I.*, *Cremonese G.*, *Da Deppo V.*, *Debei S.*, *De Cecco M.*, *Fornasier S.*, *Fulle M.*, *Groussin O.*, *Gutiérrez P.J.*, *Hviid S.F.*, *Jorda L.*, *Knollenberg J.*, *Kramm J.-R.*, *Kührt E.*, *Küppers M.*, *Lara L.M.*, *Lazzarin M.*, Lopez Moreno J.J., *Michalik H.*, *Sabau L.*, Thomas N., *Vincent J.-B.*, *Wenzel K.-P.* Comet 67P/Churyumov-Gerasimenko: Constraints on its origin from OSIRIS observations // Astron. and Astrophys. 2015. V. 583. Article id. A44. 8 p.

*Rickman H.*, *Wiśniowski T.*, *Gabryszewski R.*, *Wajer P.*, *Wójcikowsk, K.*, *Szutowicz S.*, *Valsecchi G.B.*, *Morbidelli A.* Cometary impact rates on the Moon and planets during the late heavy bombardment // Astron. and Astrophys. 2017. V. 598. Article id. A67. 15 p.

*Rotundi A, Sierks H. Della Corte V.*, *Fulle M.*, *Gutierrez P.J.*, *Lara L.*, *Barbieri C.*, *Lamy P.L.*, *Rodrigo R.*, *Koschny D.*, *Rickman H.*, *Keller H.U.*, *López-Moreno J.J.*, *Accolla M.*, *Agarwal J.*, *A'Hearn M.F.*, *Altobelli N.*, *Angrilli F.*, *Barucci M.A.*, *Bertaux J.-L.*, *Bertini I.*, *Bodewits D.*, *Bussoletti E.*, *Colangeli L.*, *Cosi M.*, *Cremonese G.*, *Crifo J.-F.*, *Da Deppo V.*, *Davidsson B.*, *Debei S.*, *De Cecco M.*, *Esposito F.*, *Ferrari M.*, *Fornasier S.*, *Giovane F.*, *Gustafson B.*, *Green S.F.*, Groussin O., *Grün E.*, *Güttler C.*, *Herranz M.L.*, *Hviid S.F.*, *Ip W.*, *Ivanovski S.*, *Jerónimo J.M.*, *Jorda L.*,

*Knollenberg J.*, *Kramm R.*, *Kührt E.*, *Küppers M.*, *Lazzarin M.*, *Leese M.R.*, Lopez- Jimenez A.C., *Lucarelli F.*, *Lowry S.C.*, *Marzari F.*, *Epifani E.M.*, *McDonnell J.A.M.*, *Mennella V.*, *Michalik H.*, *Molina A.*, *Morales R.*, *Moreno F.*, *Mottola S.*, *Naletto G.*, *Oklay N.*, *Ortiz J.L.*, *Palomba E.*, Palumbo P., *Perrin J.-M.*, *Rodríguez J.*, *Sabau L.*, *Snodgrass C.*, *Sordini R.*, *Thomas N.*, *Tubiana C.*, Vincent J.-B., *Weissman P.*, Wenzel K.-P., *Zakharov V.*, *Zarnecki J.C.* Dust measurements in the coma of comet 67P/Churyumov-Gerasimenko inbound to the Sun // Science. 2015. V. 347. Article id. aaa3905.

*Rubin M.*, *Altwegg K.*, *Balsiger H.*, *Bar-Nun A.*, *Berthelier J.-J.*, *Bieler A.*, *Bochsler P.*, *Briois C.*, *Calmonte U.*, *Combi M.*, *De Keyser J.*, *Dhooghe F.*, *Eberhard, P.*, *Fiethe B.*, *Fuselier S.A.*, *Gasc S.*, *Gombosi T.I.*, *Hansen K.C.*, *Hässig M.*, *Jäckel A.*, *Kopp E.*, *Korth A.*, *Le Roy L.*, *Mall U.*, *Marty B.*, *Mousis O.*, *Owen T.*, *Rème H.*, *Sémon T.*, *Tzou C.-Y.*, *Waite J.H.*, *Wurz P.* Molecular nitrogen in comet 67P/Churyumov-Gerasimenko indicates a low formation temperature // Science. 2015. V. 348. aaa6100.

*Schwartz S.R.*, *Michel P.*, *Jutzi M.*, *Marchi S.*, *Zhang Y.*, *Richardson D.C.* Catastrophic disruptions as the origin of bilobate comets // Nature Astronomy. 2018. V. 2, in press.

*Spohn T.*, *Knollenberg J.*, *Ball A.J.*, *Banaszkiewicz M.*, *Benkhoff J.*, *Grott M.*, *Grygorczuk J.*, *Hüttig C.*, *Hagermann A.*, *Kargl G.*, *Kaufmann E.*, *Kömle N.*, *Kührt E.*, *Kossacki K.J.*, *Marczewski W.*, *Pelivan I.*, *Schrödter R.*, *Seiferlin K.* Thermal and mechanical properties of the near-surface layers of comet 67P/Churyumov-Gerasimenko // Science. 2015. V. 349. aab0464.

*Thomas N.*, *Sierks H.*, *Barbieri C.*, *Lamy P.L.*, *Rodrigo R.*, *Rickman H.*, *Koschny D.*, *Keller H.U.*, *Agarwal J.*, *A'Hearn M.F.*, *Angrilli F.*, *Auger A.-T.*, *Barucci M.A.*, *Bertaux J.-L.*, *Bertini I.*, Besse S., *Bodewits D.*, *Cremonese G.*, *Da Deppo V.*, *Davidsson B.*, *De Cecco M.*, *Debei S.*, El-Maary M.R., *Ferri F.*, *Fornasier S.*, *Fulle M.*, *Giacomini L.*, *Groussin O.*, *Gutierrez P.J.*, *Güttler C.*, *Hviid S.F.*, *Ip W.-H.*, *Jorda L.*, *Knollenberg J.*, *Kramm J.-R.*, *Kührt E.*, *Küppers M.*, *La Forgia F.*, *Lara L.M.*, *Lazzarin M.*, *Moreno J.J.L.*, *Magrin S.*, *Marchi S.*, *Marzari F.*, *Massironi M.*, *Michalik H.*, Moissl R., *Mottola S.*, *Naletto G.*, *Oklay N.*, *Pajola M.*, *Pommerol A.*, *Preusker F.*, *Sabau L.*, *Scholten F.*, *Snodgrass C.*, *Tubiana C.*, *Vincent J.-B.*, *Wenzel K.-P.* The morphological diversity of comet 67P/Churyumov-Gerasimenko // Sience. 2015. V. 347. aaa0440.

*Vincent, J.-B.*, *Bodewits D.*, *Besse S.*, *Sierks H.*, *Barbieri C.*, *Lamy P.*, *Rodrigo R.*, *Koschny D.*, *Rickman H.*, *Keller H.U.*, *Agarwal J.*, *A'Hearn M.F.*, *Auger A.-T.*, *Barucci M.A.*, *Bertaux J.-L.*, *Bertini I.*, *Capanna C.*, *Cremonese G.*, *Da Deppo V.*, *Davidsson B.*, *Debei S.*, *De Cecco M.*, *El-Maarry M.R.*, *Ferri F.*, *Fornasier S.*, *Fulle M.*, *Gaskell R.*, *Giacomini L.*, *Groussin O.*, *Guilbert-Lepoutre A.*, *Gutierrez-Marques P.*, *Gutiérrez P.J.*, *Güttler C.*, *Hoekzema N.*, *Höfner S.*, Hviid S.F., *Ip W.-H.*, *Jorda L.*, *Knollenberg J.*, *Kovacs G.*, *Kramm R.*, *Kührt E.*,; *Küppers M.*, La Forgia F., Lara L.M., *Lazzarin M.*, *Lee V.*, *Leyrat C.*, *Lin Z.-Y.*, Lopez Moreno J.J., *Lowry S.*, *Magrin S.*, Maquet L., *Marchi S.*, *Marzari F.*, *Massironi M.*, *Michalik H.*, *Moissl R.*, *Mottola S.*, *Naletto G.*, *Oklay N.*, Pajola M., *Preusker F.*, *Scholten F.*, *Thomas N.*, *Toth I.*, *Tubiana C.* Large heterogeneities in comet 67P as revealed by active pits from sinkhole collapse // Nature. 2015. V. 523. P. 63-66.

*Vincent J.-B.*, *Hviid S.F.*, *Mottola S.*, *Kuehrt E.*, *Preusker F.*, *Scholten F.*, *Keller H.U.*, *Oklay N.*, *De Niem D.*, *Davidsson B.*, *Fulle M.*, *Pajola M.*, *Hofmann M.*, *Hu X.*, *Rickman H.*, *Lin Z.-Y.*, *Feller C.*, *Gicquel A.*, *Boudreault S.*, *Sierks H.*, *Barbieri C.*, *Lamy P.L.*, *Rodrigo R.*, *Koschny D.*, *A'Hearn M.F.*, *Barucci M.A.*, *Bertaux J.-L.*, *Bertini I.*, *Cremonese G.*, *Da Deppo V.*, *Debei S.*; *De Cecco M.*, *Deller J.*, *Fornasier S.*, *Groussin O.*, *Gutiérrez,P.J.*, *Gutiérrez-Marquez P.*, *Güttler C.*, *Ip W.-H.*, *Jorda L.*, *Knollenberg J.*, *Kovacs G.*, *Kramm J.-R.*, *Küppers M.*, *Lara L.M.*, *Lazzarin M.*, Lopez Moreno J.J., *Marzari F.*, *Naletto G.*, *Penasa L.*, *Shi X.*, *Thomas N.*, *Toth I.*, *Tubiana C.* Constraints on cometary surface evolution derived from a statistical analysis of 67P's topography // Monthly Notices of the Royal Astronomical Society. 2017. V. 469. P. S329-S338.

*Wahlberg Jansson K.*, *Johansen A.* Formation of pebble-pile planetesimals // Astron. and Astrophysics. 2014. V. 570. Article id. A47. 10 p.

*Weidenschilling S.J.* The origin of comets in the Solar nebula: A unified model // Icarus. 1997. V. 127. P. 290-306.

*Weidling R.*, *Güttler C.*, *Blum J.*, *Brauer F.* The physics of protoplanetesimal dust agglomerates. III. Compaction in multiple collisions // Astrophys. J. 2009. V. 696. P. 2036-2043.

*Whipple F.L.* A comet model. I. The acceleration of Comet Encke // Astrophys. J. 1950. V. 111. P. 375–394.

*Whizin A.D.*, *Blum J.*, *Colwell J.E.* The physics of protoplanetesimal dust agglomerates. VIII. Microgravity collisions between porous $SiO_2$ aggregates and loosely bound agglomerates // Astrophys. J. 2017. V. 836. Article id. 94. 9 p.

*Youdin A.N.*, *Goodman J.* Streaming instabilities in protoplanetary disks // Astrophys. J. 2005. V. 620. P. 459-469.

*Zellner N.E.B.* Cataclysm no more: new views on the timing and delivery of lunar impactors // Origins of Life and Evolution of Biospheres. 2017. V. 47. P. 261-280.

*Zsom A.*, *Ormel C.W.*, *Güttler C.*, *Blum J.*, *Dullemond C.P.* The outcome of protoplanetary dust growth: pebbles, boulders, or planetesimals? II. Introducing the bouncing barrier // Astron. and Astrophys. 2010. V. 513. Article id. A57. 22 p.